\documentclass[aps,prb,twocolumn,floatfix]{revtex4}
\usepackage{amsmath}
\usepackage{amssymb}
\usepackage{graphicx}
\usepackage{dcolumn}
\usepackage{bm}
\usepackage{wasysym}
\usepackage{color}
\begin{document}

\title{Quantum Hall droplet laterally coupled to a quantum ring}

\author{E. T\"ol\"o}

\author{A. Harju} 

\affiliation{Helsinki Institute of Physics and Department of Applied Physics, Helsinki University of Technology, P.O. Box 4100, FI-02015 HUT, Finland}

\begin{abstract}
We study a two-dimensional cylindrically-symmetric electron droplet separated from a surrounding electron 
ring by a tunable barrier using the exact diagonalization method. The magnetic field is 
assumed strong so that the electrons become spin-polarized and reside on the lowest 
Fock-Darwin band. We calculate the ground state phase diagram for 6 electrons. At weak coupling, 
the phase diagram exhibits a clear diamond structure due to the blockade caused by the angular
momentum difference between the two systems. We find separate excitations of the droplet and the 
ring as well as the transfer of charge between the two parts of the system. At strong coupling, 
interactions destroy the coherent structure of the phase diagram, while individual phases are still
heavily affected by the potential barrier.
\end{abstract}

\maketitle

\section{Introduction}

Fractional quantum Hall (FQH) fluids are observed at low temperature in clean two-dimensional 
electron gas exposed to a perpendicular magnetic field.\cite{tsui,laughlin,jain} 
Each of them possesses unique topological order and is characterized by a number of 
universal invariants such as the Hall conductance and the fractional charges and braiding 
statistics of the elementary quasiparticle excitations.\cite{wen_adv,nayak} 
In quantum Hall droplets realizable in semiconductor quantum dots, a microscopic 
number of electrons forms an analogous correlated fluid. While the universalities 
are no longer exactly valid, the system can be accurately studied by various
non-perturbative numerical methods combined with the understanding of the quantum Hall 
effects.\cite{henri,saarikoski_pf}

In this paper, we investigate a cylindrically-symmetric electron droplet separated 
from a surrounding electron ring by a tunable barrier. The magnetic field is assumed
strong so that the electrons are spin-polarized and reside on the lowest Fock-Darwin 
band. \cite{fock} We find the ground state phase diagram for 6 electrons. When the 
ring and dot are weakly coupled, the ground state angular momenta form a diamond structure,
following from the conservation of the total angular momentum in the combined system. This 
structure is gradually lost as the systems become strongly coupled. Finally, the effects of 
the potential wall when all the electrons are in the dot are analyzed.

An earlier theoretical study of transport properties of a similar ring-dot structure, 
albeit with weaker magnetic field, showed evidence of a transport blockade due 
to the system geometry.\cite{guclu} Later it was suggested that due to the piecewise 
linear dependence of the singlet-triplet splitting, a ring-dot geometry would be a good 
candidate for a realization of magnetic field controllable pair of spin qubits.\cite{szafran}
We are unaware of experiments conducted with this type of concentric laterally coupled 
quantum-dot-quantum-ring systems. However, in a recent experiment with coupled concentric 
quantum rings, Aharonov-Bohm periods matching the radius of both rings were found.\cite{muhle} 
Persistent current and capacitance oscillations in side coupled and embedded ring-dot systems
have been studied.\cite{buttiker}
Interference and phase phenomena in a quantum dot molecule embedded in a ring interferometer 
have been experimented,\cite{ihn} and the Fano effect in a side coupled quantum ring and dot 
has been experimentally observed.\cite{fuhrer}

The rest of the paper is organized as follows. In Section II, we present the model and the
numerical methods. Section III contains the results summarized in Section IV.

\section{Model and method}

The system is modelled by an effective-mass Hamiltonian 
\begin{equation}
\label{ham}
H=\sum^N_{i=1}\left[
 \frac{({\bf p}_i+\frac{e}{c} {\bf A}_i )^2}{2 m^*}
+ V(r_i)\right] +  \sum_{i<j}
\frac{e^2}{\epsilon r_{ij}}\ ,
\end{equation}
where $N$ is the number of electrons, and ${\bf A}$ is the vector potential
of the homogeneous  magnetic field ${\bf B}$ perpendicular to the plane.  
The material dependent parameters are $m^*=0.067m_e$, the effective mass 
of an electron, and $\epsilon=12.7$ (CGS), the dielectric constant of GaAs 
semiconductor medium. The confinement potential and the barrier between the
quantum dot and quantum ring are given by
\begin{equation}
V(r)=\frac{m^*\omega_0^2r^2}{2}+C\delta(r-r_0)\ .
\end{equation}
The potential barrier separating the two subsystems
is approximated by a delta function at radius $r_0$ scaled by strength $C$.
In this model that aims to capture the essential properties, we neglect the
effects of small thickness of the sample\cite{tolo} and screening by near-by 
metallic gates to the Coulomb interaction.

In the calculations, we set the confinement strength 
$\hbar\omega_0$ to $2\,{\rm meV}$ as its scaling should merely shift the 
ranges of magnetic fields for different phases. Lengths are written in units 
of oscillator length $l=\sqrt{\hbar/(m^*\omega)}$, where 
$\omega=\sqrt{\omega_0^2+(\omega_c/2)^2}$ and  $\omega_c=eB/(m^*c)$ is the 
cyclotron frequency. 
The ground state of Eq. (\ref{ham}) is solved by constructing the many-body 
Hamiltonian matrix in the basis of spin-polarized lowest Fock-Darwin band and 
finding its lowest eigenstate by the Lanczos diagonalization. The former 
constitutes a Landau level projection, an approximation that is valid at the 
high magnetic field regime.\cite{siljamaki}

The single-particle wave functions in oscillator units read
\begin{equation}
\label{fd}
\langle z|m\rangle=\tfrac{1}{\sqrt{\pi m!}}z^me^{-z\bar{z}/2}\ , \ \ m\geqslant0\ ,
\end{equation}
where $z=x+iy$. The non-trivial quantities are the interaction matrix elements 
$\langle m',n'|1/r_{12}|m,n\rangle$. Utilizing the angular momentum conservation 
$m'+n'=m+n$, these can be written in terms of 
\begin{equation}
M^l_{mn}=\langle m+l,n|1/r_{12}|m,n+l\rangle\ , \ \ l,m,n\geqslant0\ ,
\end{equation}
for which a particularly stable analytic formula has been derived by Tsiper. \cite{tsiper_c}

The first step in our analysis is to select the tunable parameters and calculate 
the ground state angular momentum phase diagram. For illustrational reasons, we 
content ourselves to two continuous parameters, the magnetic field $B$ and the 
position of the potential barrier $r_0$. By application of the following scaling
trick, which corresponds to curvilinear representation of the tunable parameters,
the number of Lanczos diagonalizations needed to compute a phase diagram  can be 
greatly reduced.

\begin{figure}[htb]
\includegraphics[width=0.6\columnwidth]{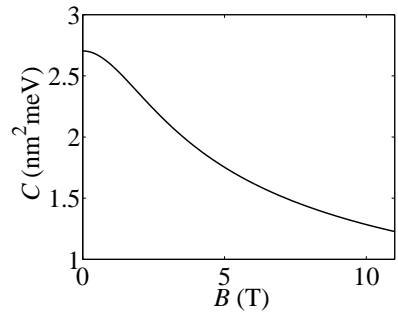}
\caption{Scaling of the potential barrier parameter $C=(l^3/a_*)\hbar\omega\beta$ with the magnetic field for $\beta=1$.}
\label{alpha0}
\end{figure}

Up to a constant shift, the energy $E$ of the ground state $|\Psi\rangle$ of the 
ring-dot system for a fixed total angular momentum $M$ reads 
\begin{equation}
E=\left(\hbar\omega-\frac{\hbar\omega_c}{2}\right)M+\hbar\omega\left[
\frac{CV_{B}}{\hbar\omega l^2}+\frac{l}{a_*}V_C\right]\ ,
\end{equation}
where $a_*=4\pi\epsilon\hbar^2/(m^*e^2)$ is the effective Bohr radius, 
$V_B=\langle\Psi|\sum_i\delta(r_i-r_0)|\Psi\rangle$ in units of $1/l^2$, and 
$V_C=\langle\Psi|\sum_{i<j}1/{r_{ij}}|\Psi\rangle$ in units of $1/l$. Hence, we 
find it convenient to scale the strength of the potential barrier as 
$C=(l^3/a_*)\hbar\omega\beta$ (see Fig.~\ref{alpha0}) with a dimensionless potential barrier strength 
parameter $\beta$, fixed for a given phase diagram. The ground state phase diagram 
is then efficiently computed by performing the exact diagonalization at each angular 
momentum value at each position (in units of $l$) of the barrier, after which the magnetic field 
dependence of the energies can be easily obtained without need for doing the 
diagonalization at each magnetic field value. After the angular momenta of adjacent 
phases are determined, the exact value of the magnetic field $B$ at the boundary is 
solved from an algebraic equation. This also means that the wave functions in a 
given phase remain the same except for scaling of $l$ as we move in the $B$-direction. 
As the energy differences of the single-particle states scale roughly as $1/B$
at high magnetic fields, the effect of the barrier is expected, despite the decrease 
of $C$, to slightly increase with increasing magnetic field.

\section{Results}
\subsection{Phase diagrams}
\begin{figure*}[htb]
\includegraphics[width=2.0\columnwidth]{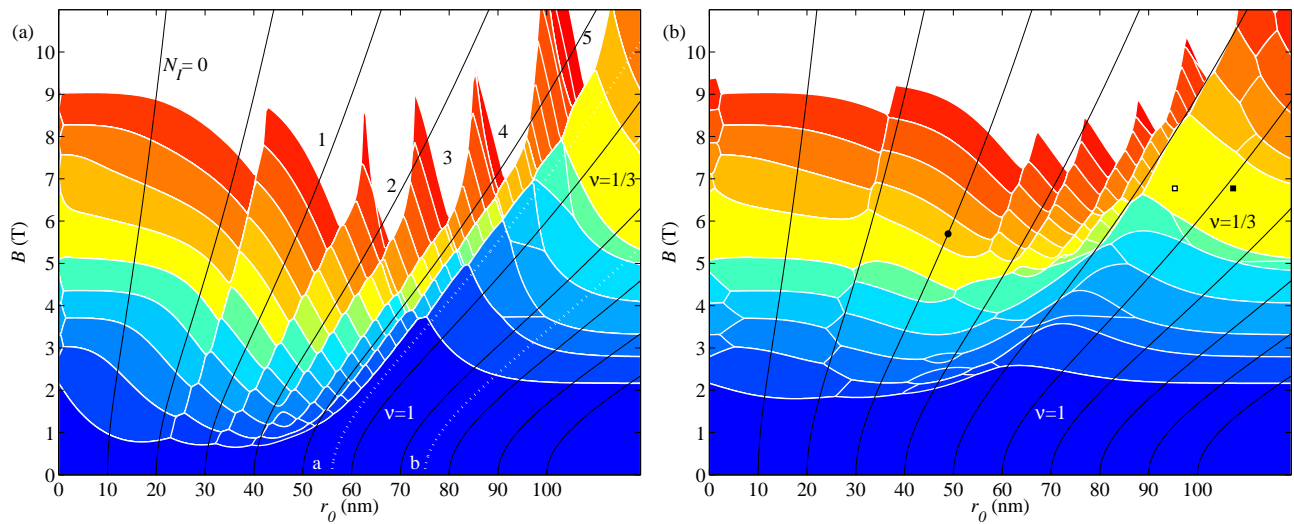}
\caption{(Color online) (a) The ground state angular momentum phase diagram for a strong barrier 
($\beta=1$) six-electron system as a function of the position of the potential 
barrier $r_0$ indicated by black lines, and the magnetic field $B$. The curvilinearity
of the co-ordinate axis is due to the reason that when we move in the vertical direction
in the phase diagram, $r_0$ in units of the oscillator length $l$ remains dimensionless 
constant as the magnetic field changes. The diagram separates into seven branches 
according to the number of electrons in the inner droplet $N_I$ indicated in the 
figure. (b) The same for a weak barrier ($\beta=0.2$). 
The color (lightness) indicates the angular momentum. For example, yellow (the lightest gray) corresponds to $M=45$, the
angular momentum for filling fraction $\nu=1/3$ state found at large $r_0$.
}
\label{faasit}
\end{figure*}

In the following, we consider the ground states of ring-dot systems with weak and 
strong coupling as a function of the magnetic field $B$ and the position of the
potential barrier $r_0$. 
The corresponding phase diagrams in the case of $N=6$ electrons are shown in 
Fig.~\ref{faasit}(a) for a weak coupling $\beta=1$ and Fig.~\ref{faasit}(b) 
for a stronger coupling $\beta=0.2$. The angular momentum $M$ increases as the magnetic
field gets stronger and most of the phases belong to a regular diamond structure  that is explained below.
\begin{figure}[htb]
\includegraphics[width=1.0\columnwidth]{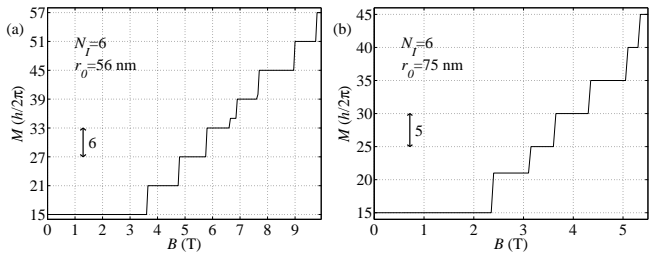}
\caption{The ground state angular momentum along lines of constant $r_0$ in the $N_I=6$ branch of
Fig. \ref{faasit}(a).}
\label{outgang}
\end{figure}

In both diagrams, the bottom phase corresponds to the minimum angular momentum of the maximum 
density droplet, in which the $N$ lowest angular momentum orbitals are compactly occupied. 
The wave function is the same as for the integer quantum Hall effect filling fraction $\nu=1$
\begin{equation}
\begin{split}
\psi(z_1,\ldots,z_N)=&\frac{e^{-\sum_{i=1}^N z_i\bar{z}_i/2}}{\sqrt{\pi^N1!\cdot2!\cdots N!}}
\begin{tabular}{|cccc|}
$1$ & $1$ & $\cdots$ & $1$\\
$z_1$ & $z_2$ & $\cdots$ & $z_N$\\
$\vdots$ & $\vdots$  & $\vdots$ &  $\vdots$ \\
$z_1^{N-1}$ & $z_2^{N-1}$ & $\cdots$ & $z_N^{N-1}$\\
\end{tabular}\\
=&\mathcal{N}\prod_{i<j}(z_i-z_j)e^{-\sum_{i=1}^N z_i\bar{z}_i/2}\ ,
\end{split}
\end{equation}
with angular momentum $M=N(N-1)/2$.

At higher angular momenta above $\nu=1$,
the potential barrier leads to the partition of the phase 
diagram according to the number of electrons $N_I=0,1,2,3,4,5,$ and $6$ in the inner 
electron droplet, or equivalently $N-N_I$ in the outer. At small $r_0$, when the 
$\delta$-function potential is close to the center of the external parabolic confinement 
potential, the density at the center vanishes and the electrons form a ring-like structure. 
When $r_0$ is increased by moving the potential barrier farther away from the origin, 
electrons tunnel into the quantum dot at the center of the system one by one. Finally, 
all electrons are in the dot, and the barrier only compresses the system giving rise to 
a reduced radius of the electron density by shifting the ground state transitions to 
higher magnetic fields and reorganizing the electron configuration. We start the detailed
analysis of the phase diagrams from this regime.

We first analyse the system following lines a and b of Fig.~\ref{faasit}(a). The corresponding 
angular momenta
are shown in Fig.~\ref{outgang}. As the magnetic field gets stronger, the successive phases tend to 
have angular momentum difference six (a) or five (b), as the electrons favour 
either a hexagonal configuration or a pentagonal configuration with one electron at the center.\cite{tolo} 
Two exceptions are the small $M=35$ phase along line a and the $M=21$ phase along line b,
which corresponds to a $\nu=1$ state with a hole at the center.
The pentagonal phases have slightly larger radius than the corresponding 
phases that support hexagonal electron configurations so that the compression eventually 
favours the hexagonal configuation and counter-intuitively actually reduces the density near the center. 
A notable exception is the strong yellow (lightest gray) phase continuously connected to the 
Laughlin state at filling fraction $\nu=1/3$ with the wave function 
\begin{equation}
\psi(z_1,\ldots,z_N)=\mathcal{N}\prod_{i<j}(z_i-z_j)^3e^{-\sum_{i=1}^Nz_i\bar{z}_i/2}\ 
\end{equation}
and angular momentum $M=3N(N-1)/2$ that supports both the pentagonal and hexagonal electron
configurations.

\subsection{Weak coupling}

Let us now analyze more carefully each branch in the weak coupling phase diagram. For this purpose, we define the
angular momentum of the dot as $M_D=\sum_m n_mm$, where the sum is taken up to such $m$ that 
$\sum_mn_m=N_I$. If this orbital is shared between the ring and the dot, only the corresponding 
fraction of $n_m$ is employed in place of $n_m$. The angular momentum of the ring is then obtained
from the total angular momentum as $M_R=M-M_D$. Note that in the coupled system $M_R$ and $M_D$ are
in general not exact integers since the wave function typically contains superpositions of the two systems
with different angular momenta summing to a given total angular momentum. 

\begin{figure*}[htb]
\includegraphics[width=2.0\columnwidth]{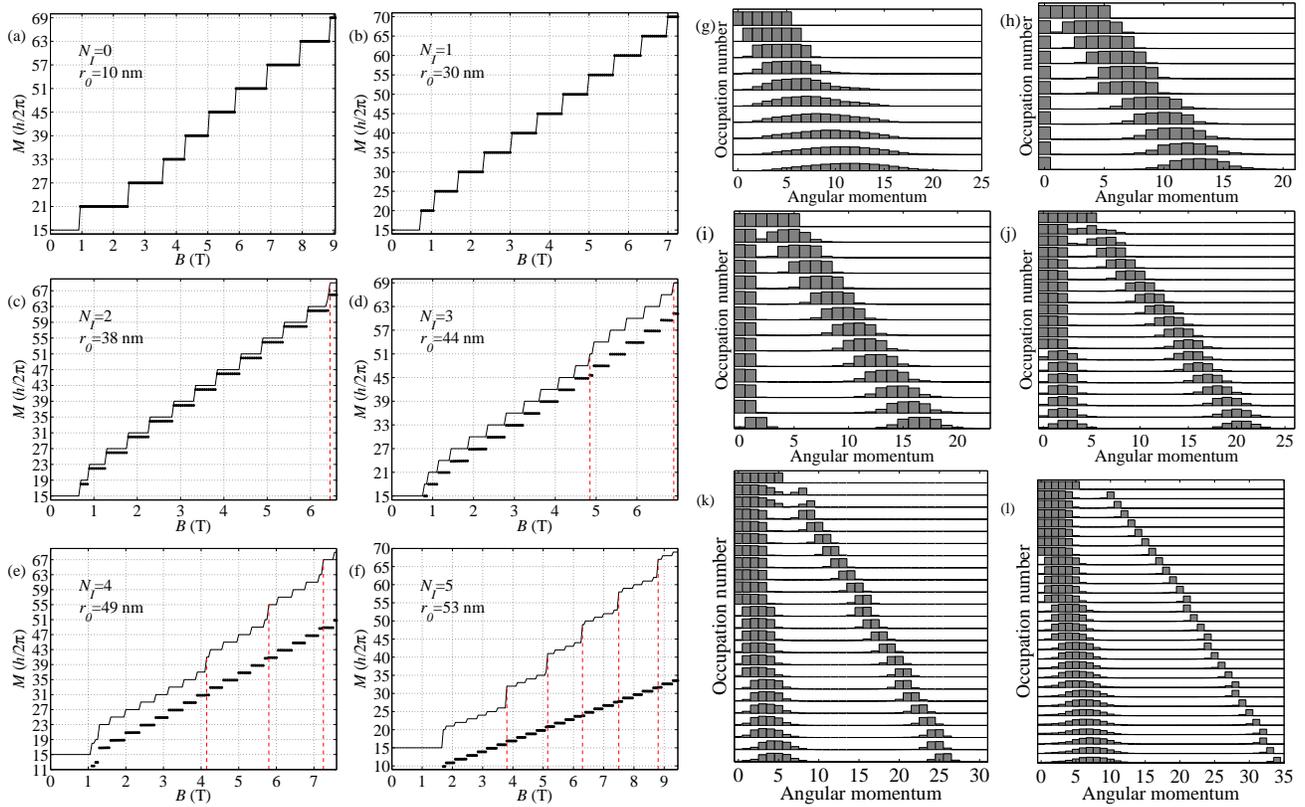}
\caption{(Color online) The total ground state angular momentum (solid-line) and the angular momentum 
of the ring (black dots) along lines of constant $r_0$ in the $N_I=0,1,2,3,4$ and $5$ branches of 
Fig. \ref{faasit}(a). The vertical dashed red lines mark the transitions of the inner system. 
Occupation numbers for the ground states in panels (a-f) evaluated at the first magnetic field
value at each plateau, the uppermost corresponding to $B=0$ T.}
\label{weakang}
\end{figure*}

Figs.~\ref{weakang}(a-f) show
$M$ and $M_R$ as a function of $B$ along lines of constant $r_0$ chosen from branches $N_I=0$ to 
$N_I=5$ of the phase diagram.  The corresponding occupation numbers for each ground-state angular 
momentum at the first magnetic field value on each plateau are listed in Figs.~\ref{weakang}(g-l),
where number of electrons in the dot $N_I$ run from $0$ to $5$.
In Figs.~\ref{weakang}(a) and (g) the angular momentum difference $\Delta M$ is always 6 indicating that the electrons form a 
hexagonal ring. Next in Figs.~\ref{weakang}(b) and (h), $\Delta M=5$ and $M=M_R$ so that one electron occupies the $m=0$ orbital
and the remaining five electrons are in the ring. Fig.~\ref{weakang}(c) and (i) tell similar story, where now two first orbitals
are occupied while the remaining 4 electrons are in the ring yielding $\Delta M=4$ and $M-M_R=1$. 
At about 6.5 T, there is a transition in the inner system as two electrons at the center excite to 
$m=1$ and $m=2$ giving $M-M_R=3$. In Fig.~\ref{weakang}(d) and (j), there are two such transitions as $M-M_R$ is first 3, 
then it increases to 6 and finally to 9. In Figs.~\ref{weakang}(e,k) and (f,l), we have similarly the frequent excitations 
of the ring with $\Delta M=2$ and $1$, respectively, and the less frequent excitations of the dot
with $\Delta M=4$ and $5$, respectively. In the occupation numbers (Figs.~\ref{weakang}(g-l)), these transition are 
seen such that subsequent states differ either by the inner or outer (or both) system moving one unit to the right.

We thus see that for $N_I<6$ the premier mechanism for gaining angular momentum is to periodically 
give the incremental angular momentum to the electrons of the exterior ring, which 
leads to the angular momentum difference $\Delta M=N-N_I$ for successive phases. The 
magnetic field range of each phase is then roughly a constant. 
The subsidiary mechanism is to increase the angular momentum of the inner system, 
which typically involves creation of a vortex near the center. In this case, the angular
momentum shift is around $N$ depending on whether all or merely the inner electrons move
radially outwards. 

The transfer of electrons in and out of the quantum ring should be observable in the
Aharonov-Bohm magnetic period.\cite{szafran} We also expect the creation of vortices 
in the electron droplet to be observable in the conductivity in analogy with findings of 
Ref. \onlinecite{muhle} for concentric quantum rings. In fact, by comparing the discs defined 
by the radius of density maximum in the quantum dot between the vertical dashed red lines 
in Fig. \ref{weakang}, we found that the area of the disc increased approximately by one 
in oscillator units at each transition, which in the limit $\omega_0\rightarrow0$ 
corresponds to one magnetic flux quantum. Similar behaviour is observed for the area enclosed 
by the density maximum of the ring.

\begin{figure*}[thb]
\includegraphics[width=2.0\columnwidth]{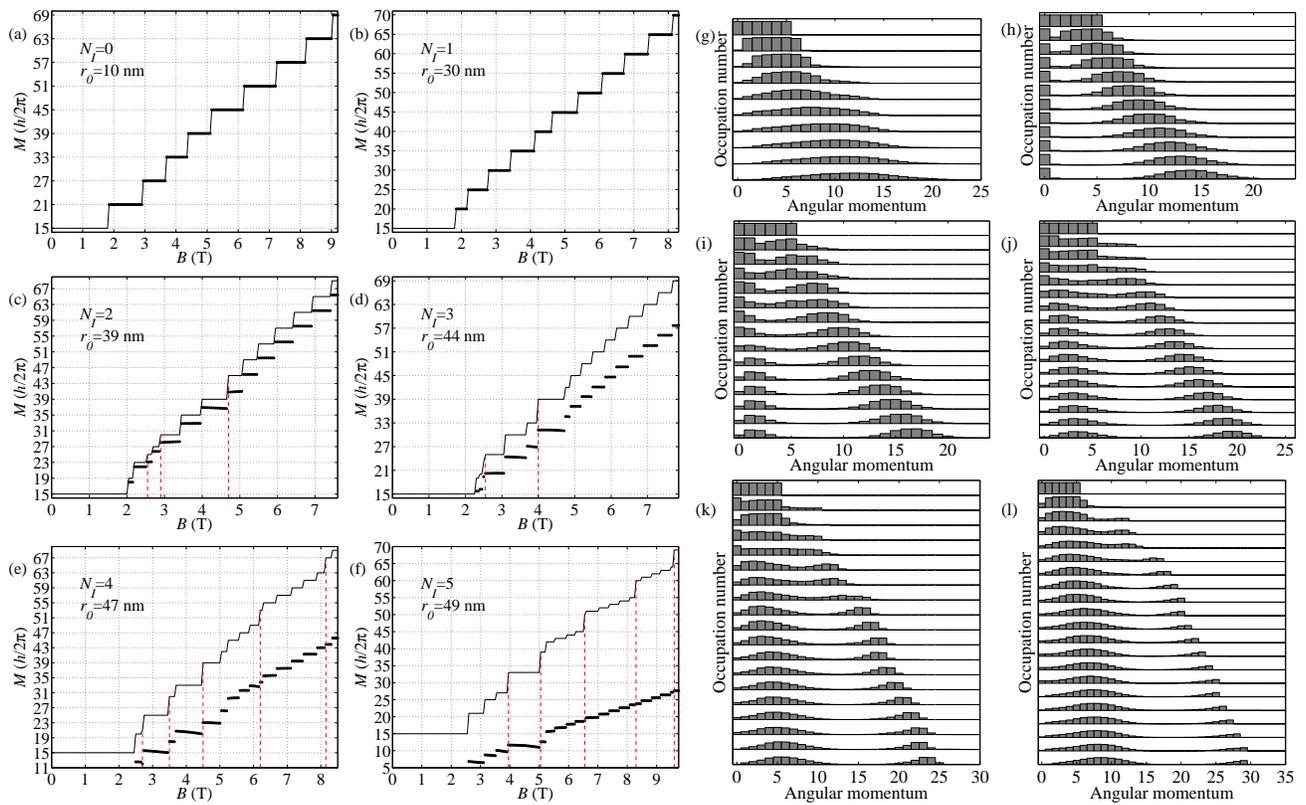}
\caption{(Color online) The same as Fig.~\ref{strongang} for $\beta=0.2$.}
\label{strongang}
\end{figure*}

The regular diamond structure in Fig. \ref{faasit}(a) is due to the symmetry blockade caused 
by the fact that the transfer of charge in and out of the droplet constitutes an immense
change in the angular momentum as a large number of radially localized angular momentum 
orbitals fit in the space between the ring and the dot. We remark that the phases tend to get thinner 
in $r_0$ direction as we move to larger $r_0$ 
because the same amount of angular momentum orbitals become available for a smaller increase of $r_0$. 
This is due to the fact that the radial size of the orbital with quantum number $m$ is 
proportional to $\sqrt{m}$ rather than $m$.

The small phases near $r_0=0$, more visible for the weaker barrier in Fig. \ref{faasit}(b), are 
due to the fact that a normal parabolic quantum dot with six electrons supports the pentagonal electron 
configurations except at $M=21$ and $M=39$.\cite{tolo} Since the influence of the potential is proportional 
to its perimeter, at sufficiently small $r_0$ the potential can no longer force the hexagonal
configuration, and the ground-state angular momenta of a parabolic quantum dot, found also at the limit 
$r_0\rightarrow\infty$, are reacquired.

\subsection{Strong coupling}

Recall the ground state angular momentum phase diagram for $\beta=0.2$ shown in Fig.~\ref{faasit}(b). 
As the strength of the barrier $\beta$ is weakened, the regular structure of the phase diagram present 
at a strong barrier strength is gradually lost as the coupling between the inner and outer system 
becomes larger and a more unilateral collective behaviour extends over the barrier. Due to scaling of 
the energies with the magnetic field as mentioned in Sec.~II, the coherent structure fails first at 
weak magnetic fields.

Figs.~\ref{strongang}(a-f) show $M$ and $M_R$ as a function of $B$ along lines of constant $r_0$ 
chosen from branches $N_I=0$ to $N_I=5$ of the phase diagram with the occupation numbers at each
plateau listed in Figs.~\ref{strongang}(g-l). 
Figs.~\ref{strongang}(a,g) and (b,h) admit to the
explanation of hexagonal and pentagonal electron configurations with 0 and 1 electron at the center
as previously. At strong magnetic fields, Figs.~\ref{strongang}(c-f) and (i-l) are reminiscent of the weak coupling case with
excitations of the dot and the angular momentum of the ring increasing in definite steps. In the higher 
angular momentum states, the
separation into two systems is evident in the occupation numbers.
Main difference to the strong barrier case is the spreading of the angular momentum reflecting 
the strong coupling of the systems, best seen in Fig.~\ref{strongang}(l). However, at
weak magnetic fields the ring angular momentum is no longer constant suggesting that our simple
picture breaks down due to the electron correlations. This is seen on the first few lines of the 
corresponding occupations in Figs.~\ref{strongang}(j-l) where the electrons appear to be either in the dot or an excited dot.

In Fig.~\ref{strongang}(i), the three small phases crossed by $r_0=40$ nm in 
Fig.~\ref{faasit}(b) still have large weight on the two first orbitals corresponding to 2 electrons at 
the center and the quantity $N-\sum_{m=0}^{N-1}n_m$, 
where $n_m$ is the occupation of angular momentum orbital $m$, is exceedingly close to 1,2, and 3, 
respectively, demonstrating quantized charge depletion relative to the $\nu=1$ maximum density 
droplet. For the phases directly above $\nu=1$, the charge depletion is one electron charge to 
an accuracy of 1\permil.

\subsection{Edge green's function}

Even when all the electrons are inside the barrier, the compression still affects the edge 
details of the system. 
A quantity of interest for the tunnelling experiments that probe the chiral Luttinger 
liquid theory of the FQH edge in macroscopic samples
is the current-voltage power-law dependence $I\propto V^{\alpha}$, and hence (in the thermodynamic 
limit) the power-law dependence of the edge Green's function \cite{chang} 
\begin{equation}
G_e(z,z_0)=\langle\psi^{\dagger}(z)\psi(z_0)\rangle\ ,
\end{equation}
which is called the edge Green's function as it is evaluated along the edge. For an electron
disc, we may set $z_0=r$ and $z=re^{i\theta}$, where $r$ is about the radius of the disc. According 
to the theory $|G_e|\propto |z-z_0|^{-\alpha}\propto|\sin(\theta/2)|^{-\alpha}$, 
where $\alpha$ reflects the topological order of the FQH phase. The experimental results 
are somewhat puzzling in that while there appears to be a clear non-Ohmic current-voltage 
power-law dependence $I\propto V^{\alpha}$ in accordance with the chiral Luttinger liquid 
theory of the FQH edge \cite{chang,wen_adv}, the exact value of the tunnelling exponent 
$\alpha$ appears to be sample dependent and in general notably less than the supposed 
universal value derived from the bulk effective theory. A number of mechanisms, including 
taking into account the long-range nature of the Coulomb interaction and edge reconstruction, 
whose effect amounts to renormalization of the power-law exponent $\alpha$, have been suggested 
by several authors \cite{chang}. However, no clear consensus of the correct picture has been 
achieved so far.

\begin{figure}[htb]
\includegraphics[width=\columnwidth]{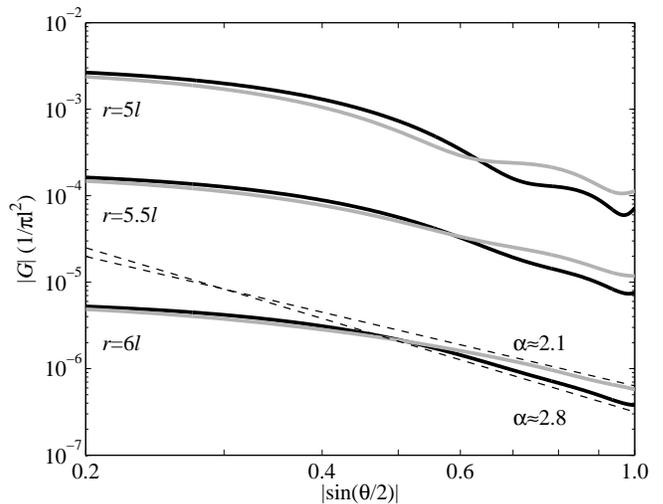}
\caption{The edge Green's functions at three radial distances $r$ for the two states in the $\nu=1/3$ topological phase marked by the squares in Fig.~\ref{faasit}(b). The black square corresponds to the black line.}
\label{greenfff}
\end{figure}

\begin{figure}[htb]
\includegraphics[width=\columnwidth]{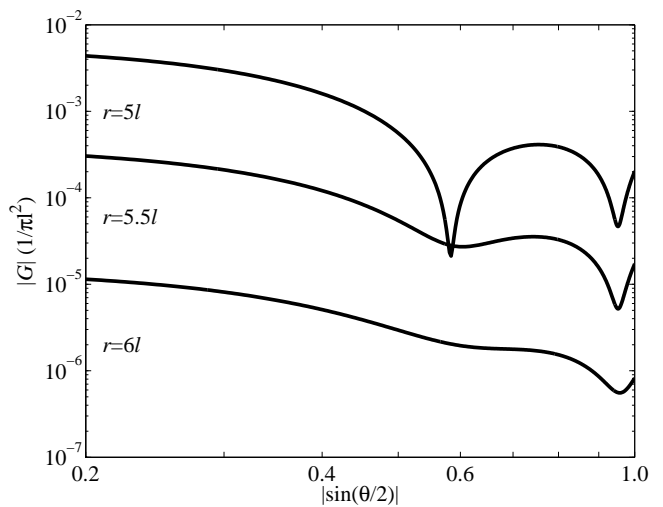}
\caption{The edge Green's function at three radial distances $r$ for at the point $r_0=30$ nm, $B=5.7$ T, and $\beta=0.2$ 
marked in the strong coupling phase diagram of Fig.~\ref{faasit}(b).}
\label{greenfff3}
\end{figure}

For the $\nu=1/3$ phase, in which it is possible to continuously interpolate the 
superposition of the hexagonal and pentagonal electron configurations, the tunnelling exponent 
$\alpha$ decreases continuously as the compression is increased. This result is in agreement 
with the findings of Ref.~\onlinecite{wan}, where with a slightly different confinement potential and 8 electrons
the exponent $\alpha$ was found to change after cutting the size of the single-particle 
basis. Fig.~\ref{greenfff} illustrates the determination of the power-law exponent 
$\alpha$. The oscillations are seen to vanish as the edge is approached and the exponents $\alpha\approx2.1$ 
and $2.8$ can be credibly extracted. However, this appears not to be the case for many of 
the phases below and above $\nu=1/3$ as well as other parts of the phase diagram where heavy oscillations 
of the edge Green's function render 
the extraction of $\alpha$ rather uncertain (see Fig.~\ref{greenfff3}). An exception is the $\nu=1$ phase, which has the
power-law behaviour $\alpha=1$ typical of Fermi liquids.


\section{Summary}

We have studied cylindrically-symmetric electron droplet tunably coupled to a surrounding quantum ring
in the strong magnetic field regime. The structure of the ground state angular momentum phase diagram 
was understood in terms of radial charge transfer, which should be observable in a lateral transport 
Aharonov-Bohm interferometer experiment. 
The compression induced by the potential barrier was also found
to renormalize the tunnelling exponent of the edge Green's function.

\section{Acknowledgements}

This study has been supported by the Academy of Finland through its Centres of Excellence Program (2006-2011). 
ET acknowledges financial support from the Vilho, Yrj\"o, and Kalle V\"ais\"al\"a Foundation of the Finnish
Academy of Science and Letters. We also thank C. Webb for useful discussions.


\begin{thebibliography}{10}
\bibitem{fock} This corresponds to the lowest Landau level in the limit that 
the harmonic confinement potential of the quantum dot vanishes.
\end{thebibliography}
\end{document}